\documentclass{cjpsuppl}
\usepackage{epsf}
\begin{document}
\title{Diffractive hadron leptoproduction and SPD}
\authori{S.\,V. Goloskokov}
\addressi{Bogoliubov Laboratory of Theoretical  Physics,
  Joint Institute for Nuclear Research}
\authorii{}    \addressii{}
\authoriii{}   \addressiii{}
\authoriv{}    \addressiv{}
\authorv{}     \addressv{}
\authorvi{}    \addressvi{}
\headtitle{Diffractive hadron leptoproduction and SPD \ldots}
\headauthor{S.\,V. Goloskokov} \lastevenhead{S.\,V. Goloskokov:
Diffractive hadron leptoproduction and SPD} 
\refnum{}
\daterec{20 October 2002;\\final version 31 December 2003}
\suppl{A}  \year{2003} \setcounter{page}{1}
\maketitle

\begin{abstract}
We discuss diffractive hadron leptoproduction in terms of skewed
gluon distributions and within two-gluon exchange model.
Connection of the two-gluon approach with skewed gluon
distributions is found. The double spin asymmetries for
longitudinally polarized leptons and transversely polarized
protons in diffractive vector meson and $Q \bar Q$ production at
high energies  within the two-gluon model is analysed.
 The asymmetry predicted for meson
production is found to be quite small. The $A_{lT}$ asymmetry for
$Q \bar Q$ production is large and can be used to  obtain
information on the polarized skewed gluon distributions in the
proton.
\end{abstract}

\section{Introduction}
In this lecture, we study diffractive hadron leptoproduction at
high energies and small $x$.  The scattering amplitude in this
region is predominated by the the two-gluon exchange which can be
associated with the Pomeron. In diffractive reactions like vector
meson and $Q \bar Q$ production the  momentum $x_P$ carried by the
two-gluon system is nonzero and the gluon momenta cannot be equal.
Such processes can be expressed in terms of generalized or skewed
parton distribution (SPD) in the nucleon ${\cal F}_\zeta(x)$ where
$x$ is a fraction of the proton momentum carried by the outgoing
gluon and $\zeta$ is the difference between the gluon momenta
(skewedness) \cite{rad-j}. For the processes which include light
quarks, the $q \bar q$ exchange in the $t$ -channel  should be
important for $x \geq 0.1$ in addition to the gluon contribution.
For  diffractive $J/\Psi$ and charm quark production the
predominated contribution is determined by the two-gluon exchange
(gluon SPD). We shall mainly discuss here such reactions which
should play a key role in future study of the gluon distribution
${\cal F}_\zeta (x)$ at small $x$.

Intensive experimental study of diffractive processes was
performed in DESY (see e.g. \cite{zeus97,jpsi1,h1_99,dijet}). The
polarized observables in vector meson production were analysed in
\cite{dm,hermes}. Theoretical investigation of the diffractive
vector meson production was conducted on the basis of the two-
gluon exchange model where the typical scale $\bar
Q^2=(Q^2+M_V^2)/4$ was found for vector mesons production
\cite{rys,J-Psi}. The cross sections of light and heavy meson
production plotted versus this variable look similar
\cite{clebaux}.  The amplitudes for longitudinally and
transversely polarized photon were analyzed in
\cite{rys1,cud,ivan}. The longitudinally polarized photon gives a
predominant contribution to the cross section for $Q^2 \to
\infty$. The cross section with transverse photon polarization is
suppressed as a power of $Q$. Note that in the two-gluon model the
imaginary part of the amplitude is usually calculated. The real
part can be reproduced \cite{J-Psi} using the local dispersion
relations. A more general SPD approach was applied to study the
vector meson production  by many authors (see e.g
\cite{mank98,hua_kr}). Within the SPS approach, one can study
simultaneously the imaginary and real parts of diffractive
amplitudes. The double spin asymmetry for longitudinal photon and
proton polarization in $J/\Psi$ production was investigated in
\cite{mank}. Theoretical analysis of the diffractive $Q \bar Q$
production was carried out in \cite{die95, bart96, ryskin97,
schaef}. It was shown that the cross sections of diffractive
quark- antiquark production were expressed in terms of the same
gluon distributions as in the case of vector meson production.

This lecture is organized as follows. In the second section, we
discuss the main features of the diffractive hadron
leptoproduction processes and their connection with SPD. The
two-gluon model approach and the spin structure of the two-gluon
coupling with the proton are considered in section 3. The
diffractive $J/\Psi$ and charm quark pair production is analysed
in sections 4, 5. The connection of the two-gluon approach with
skewed gluon distributions is found here too. In sections 6 and 7
we consider double spin asymmetries for polarized leptons and
protons in diffractive vector meson and $Q \bar Q$ production at
high energies.   The predictions for the HERMES and COMPASS
energies are made here.

\section{Diffractive hadron production and SPD}

Let us study the diffractive hadron production in lepton-proton
reactions
\begin{equation}
\label{react} l+p \to l+p +H
\end{equation}
at high energies in a lepton-proton system. The hadron state $H$
in this reaction can contain a vector meson or a $Q \bar Q$ system
which are observed as two final jets. The reaction (\ref{react})
can be described in terms of the kinematic variables which are
defined as follows:
\begin{eqnarray}
\label{momen} q^2= (l-l')^2=-Q^2,\;t=r_P^2=(p-p')^2, \nonumber
\\  y=\frac{p \cdot q}{l  \cdot p},\;x=\frac{Q^2}{2p  \cdot q},\;
x_P=\frac{q \cdot (p-p')}{q \cdot p},\; \beta=\frac{x}{x_P},
\end{eqnarray}
where $l, l'$ and $p, p'$ are the initial and final lepton and
proton momenta, respectively, $Q^2$ is the photon virtuality, and
$r_P$ is the momentum carried by the the two-gluons (Pomeron). The
variable $\beta$ is used in $Q \bar Q$ production. In this case
the effective mass of a produced quark system is equal to
$M_X^2=(q+r_P)^2$ and can be quite large. The  variable
$\beta=x/x_P \sim Q^2/(M_X^2+Q^2)$ can vary here from 0 to 1. For
diffractive vector meson production, $M_X^2=M_V^2$ and $\beta \sim
1$ for large $Q^2$. From the mass-shell equation for the
vector--meson momentum $V^2=(q+r_P)^2=M_V^2$ we find that for
these reactions
\begin{equation}
\label{x_P} x_P \sim \frac{m_V^2+Q^2+|t|}{s y}
\end{equation}
and is small at high energies. This variable is not fixed for $Q
\bar Q$ production.

 We shall use the center of mass system of the photon and proton.
The spin-average and spin dependent cross sections with parallel
and antiparallel longitudinal polarization of a lepton and a
proton are determined by
\begin{equation}
\label{spm} \sigma(\pm) =\frac{1}{2} \left( \sigma(^{\rightarrow}
_{\Leftarrow}) \pm \sigma(^{\rightarrow} _{\Rightarrow})\right).
\end{equation}
They can be expressed in terms of the vector meson photoproduction
helicity amplitude $M_{\mu'\lambda',\mu\lambda}$, where $\mu$ and
$\lambda$ are the photon and initial proton polarization, $\mu'$
and $\lambda'$ are the vector meson  and final proton
polarization.

For the cross section integrated over the azimuthal angle between
lepton and hadron scattering plane we have
\begin{equation}
\label{s+}
  \sigma(+)=N\,\left[ \sum_{\mu',\lambda',\lambda}\,| M_{\mu'\,\lambda',+1\,\lambda}|^2
  + \epsilon \sum_{\mu',\lambda',\lambda}\,|M_{\mu'\,\lambda',0\,\lambda}|^2 \right ]
\end{equation}
and
\begin{equation}\label{s-}
\sigma(-)= - N \, \sqrt{1-\epsilon^2}\, \sum_{\mu',\lambda'}
\left(| M_{\mu'\,\lambda',+1\,\frac{1}{2}}|^2 -
|M_{\mu'\,\lambda',+1\,-\frac{1}{2}}|^2 \right ).
\end{equation}
Here $\epsilon \simeq 2 (1-y)/(1+(1-y)^2)$ is a virtual photon
polarization.

The leading contribution to the spin-average cross section can be
estimated as
\begin{equation}\label{slp}
  \sigma(+)\sim | M_{+1\,\frac{1}{2},+1\,\frac{1}{2}}|^2
  + \epsilon |M_{0\,\frac{1}{2},0\,\frac{1}{2}}|^2.
\end{equation}
The photons with longitudinal and transverse polarization
contribute here.

On the other hand, we find that only transversely polarized
photons contribute to the spin-dependent cross section
\begin{equation}
\sigma(-)\sim  \,  \sqrt{1-\epsilon^2}\,
 \left(| M_{+1\,\frac{1}{2},+1\,\frac{1}{2}}|^2
 -|M_{+1\,-\frac{1}{2},+1\,-\frac{1}{2}}|^2 \right).
\end{equation}
The model analysis of the amplitudes with different photon and
vector meson polarization shows that \cite{ivan}
\begin{equation}
\frac{|M_{+1\,\frac{1}{2},+1\,\frac{1}{2}}|}{|M_{0\,\frac{1}{2},0\,\frac{1}{2}}|}
\sim 0.6; \;
\frac{|M_{0\,\frac{1}{2},+1\,\frac{1}{2}}|}{|M_{0\,\frac{1}{2},0\,\frac{1}{2}}|}
\sim 0.12; \;
\frac{|M_{+1\,\frac{1}{2},0\,\frac{1}{2}}|}{|M_{0\,\frac{1}{2},0\,\frac{1}{2}}|}
\sim 0.06.
\end{equation}
Such amplitude hierarchy does not contradict the experimental data
on the polarized density matrix elements \cite{dm} of the vector
meson production. Thus, we can conclude that the amplitude with
transversally polarized photons is very important in spin
observables. Unfortunately, for the light meson production, these
amplitudes are determined by the higher twist effects and are not
well defined because of the present infrared singularities
\cite{mank00}.

The photoproduction amplitude of longitudinally polarized vector
meson can be written in the factorized form \cite{rad-j}
\begin{equation}\label{ma}
M_{0\,\lambda',\mu\,\lambda}\propto \int \frac{d X}{X (X-\zeta- i
\epsilon)}\,\langle
\lambda'|F_{\rho1,\rho2}(X,\zeta,t,...)|\lambda\rangle \,
H^{\rho1,\rho2}_{0,\mu}(Q,k1,k2,...).
\end{equation}
Here $F(X,\zeta,...)$ is the soft part which can be determined in
terms of skewed parton distributions (SPD) and the hard part
$H(X,...)$ can be calculated perturbatively.

We can insert the $ g^{\rho\,\rho'}$ tensor in the intermediate
state which can be decomposed in the axial gauge as
\begin{equation}
 g^{\rho\,\rho'}=\sum_\nu \epsilon^{\rho}(k,\nu)
 \epsilon^{*\rho'}(k,\nu).
\end{equation}
Here gluons are transversally polarized.  The hard scattering
amplitudes are now determined by the relation
\begin{equation}\label{h}
H_{0\,\nu',\mu\,\nu}=\epsilon^{*}_{\rho1}(k2,\nu')\,
H^{\rho1,\rho2}_{0,\mu}\,\epsilon_{\rho2}(k1,\nu).
\end{equation}

We suppose that the gluon helicity flip is suppressed and
$\nu=\nu'$.  In this case we can write \cite{hua_kr}
\begin{equation}
\epsilon^{*\rho2}(k1,\nu)\,\epsilon^{\rho1}(k2,\nu)=\frac{1}{2}\,(g^{\rho1\,\rho2}_\perp+
\nu \,{\cal P}^{\rho1\,\rho2}).
\end{equation}
 The symmetric and asymmetric parts of gluon SPD
(\ref{gd}) can be decomposed into the following structures:
\begin{eqnarray}\label{gd}
&&\langle \lambda'|F_{\rho1,\rho2}|\lambda\rangle
\,g^{\rho1\,\rho2}_\perp= \frac{ \langle \lambda'| \hat
q'|\lambda\rangle}{p \cdot q'}\, {\cal F}^g_{\zeta}(X,t)
           + \frac{i}{2m} \frac{\langle \lambda'|\sigma_{\mu\nu}| \lambda\rangle \,q'^\mu r^\nu }
              {p \cdot q'}\, {\cal K}^g_{\zeta}(X,t)\, \nonumber\\
&&\langle \lambda'|F_{\rho1,\rho2}|\lambda\rangle \,{\cal
P}^{\rho1\,\rho2}= \frac{\langle \lambda'| \hat
q'\gamma_5|\lambda\rangle }{p \cdot q'}\,  {\cal
G}^g_{\zeta}(X,t).
\end{eqnarray}

For zero momentum transfer one can find that \cite{rad-j}
\begin{eqnarray}
M_{0\,\frac{1}{2},0\,\frac{1}{2}}=N \frac{1}{Q}\,  \frac{ \langle
\lambda'| \hat q'|\lambda\rangle}{p \cdot q'}\, f_V \int_0^1 d\tau
\frac{\Phi_V(\tau)}{\tau (1-\tau)} \int \frac{d X{\cal
F}^g_{\zeta}(X,t)}{X (X-\zeta- i \epsilon)}.
\end{eqnarray}
The Im part of the amplitude is predominated at high energies and
has the form
\begin{equation}
M_{0\,\frac{1}{2},0\,\frac{1}{2}} \sim {\rm
Im}M_{0\,\frac{1}{2},0\,\frac{1}{2}}\propto H_0 \frac{{\cal
F}^g_{\zeta}(\zeta,t)}{\zeta}.
\end{equation}
Here $H_0=H_{0\,+,0\,+}$ is the amplitude with positive gluon
polarization.

The cross sections for nonzero $|t|$ looks like
\begin{equation}
\label{s+i} \sigma(+)\sim H_0^2 \left[ {\cal
F}^g_{\zeta}(\zeta,t)^2 + \frac{|t|}{m^2} {\cal
K}^g_{\zeta}(\zeta,t)^2 \right].
   \end{equation}
The spin-dependent cross section for a longitudinally polarized
target  is expressed through ${\cal G}^g_{\zeta}(X,t)$.
\section{Two-gluon exchange model}\label{sect3}
Now let us study  the process of the hadron leptoproduction within
the two-gluon exchange model. As we have discussed previously,
this contribution is predominated at small $x \leq 0.1$.  For
larger $x$  the quark exchange should be included for processes
with light quarks. The cross section of hadron leptoproduction can
be decomposed into the leptonic and hadronic tensors and the
amplitude of hadron production through the $\gamma^\star gg$
transition to the vector meson or $Q \bar Q$ states. To study spin
effects in diffractive hadron production, one must know the
structure of the two-gluon coupling with the proton at small $x$.

It has been shown in \cite{grib77} that the leading contribution
like $\alpha_s \left[ \alpha_s \ln{(1/x)}\right]^n$ to the Pomeron
is determined by the gluon ladder graphs. The two-gluon coupling
with the proton can be parametrized in the form \cite{golostr}
\begin{eqnarray}\label{ver}
V_{pgg}^{\alpha\beta}(p,t,x_P,l_\perp)&=& B(t,x_P,l_\perp)
(\gamma^{\alpha} p^{\beta} + \gamma^{\beta} p^{\alpha}) \nonumber\\
&+&\frac{i K(t,x_P,l_\perp)}{2 m} (p^{\alpha} \sigma ^{\beta
\gamma} r_{\gamma} +p^{\beta} \sigma ^{\alpha \gamma}
r_{\gamma})+...  .
\end{eqnarray}
Here $m$ is the proton mass. In the matrix structure (\ref{ver})
we wrote only the terms with the maximal powers of a large proton
momentum $p$ which are symmetric in the gluon indices
$\alpha,\beta$. The structure proportional to $B(t,...)$
determines the spin-non-flip contribution. The term $\propto
K(t,...)$ leads to the transverse spin-flip at the vertex. If one
considers the longitudinal spin effects, the asymmetric structure
$\propto \gamma_\rho \gamma_5$ should be included in (\ref{ver}).

In what follows, we analyze the $\gamma^* g g \to Q \bar Q$
transition amplitude. The typical momentum of quarks is
proportional to the photon momentum $q$. In the Feynman gauge, we
can decompose the $g_{\mu\nu}$ tensors from $t$- channel gluons
into the longitudinal and transverse parts \cite{grib77}
\begin{equation}\label{gmunu}
g^{\alpha \alpha'}= g^{\alpha \alpha'}_l+g^{\alpha
\alpha'}_\perp\; \mbox{with}\; g^{\alpha \alpha'}_l\sim
\frac{q^\alpha p^{\alpha'}}{(pq)}.
\end{equation}
The product of the $g^{\alpha \alpha'}_l$ tensors and the
two-gluon coupling of the proton can be written in the form
\begin{equation}\label{g_vec}
g^{\alpha' \alpha}_l g^{\beta
'\beta}_lV_{pgg}^{\alpha\beta}(p,t,x_P,l_\perp)\propto
p^{\alpha'}p^{\beta'} \left[\frac{ /\hspace{-1.7mm} q}{p \cdot q}
B(t,x_P,l_\perp) +\frac{i K(t,x_P,l_\perp)}{2 m \; p \cdot q}
\sigma ^{\beta \gamma} q_{\beta} r_{\gamma} \right].
\end{equation}
It can be seen that the structure in the square brackets in
(\ref{g_vec}) after integration over the gluon momentum $l_\perp$
should be related  directly to the skewed gluon distribution
(\ref{gd})
\begin{equation}\label{spd}
{\cal F}^g_\zeta (\zeta,t) \propto \int d^2l_\perp
B(t,\zeta=x_P,l_\perp) \phi(l_\perp,...).
\end{equation}
A similar equation should be valid for the $K$ function. The
function $\phi$ will be found later.

In what follows we will calculate directly the spin dependent
cross section of hadron production which can be expressed in terms
of the hard hadron production amplitude through photon-gluon
fusion convoluted with the spin-dependent hadron and lepton
tensors. The structure of the leptonic tensor is simple because
the lepton is a point-like object
\begin{eqnarray}
\label{lept} {\cal L}^{\mu \nu}(s_l)&=& \sum_{spin \ s_f} \bar
u(l',s_f) \gamma^{\mu} u(l,s_l) \bar u(l,s_l) \gamma^{\nu}
u(l',s_f)
\nonumber \\
&=& {\rm Tr} \left [  (/\hspace{-1.7mm} l+\mu) \frac{1+ \gamma_5
/\hspace{-2.1mm} s_l}{2} \gamma^{\nu} ( /\hspace{-1.7mm} l'+\mu)
\gamma^{\mu}
 \right],
\end{eqnarray}
where $l$ and $l'$ are the initial and final lepton momenta, and
$s_l$ is a spin vector of the initial lepton.

Spin-average and spin--dependent  lepton tensors are defined by
\begin{equation}
\label{l+-} {\cal L}^{\mu \nu}(\pm)=\frac{1}{2}({\cal L}^{\mu
\nu}(+\frac{1}{2}) \pm {\cal L}^{\mu \nu}(-\frac{1}{2})).
\end{equation}
The ${\cal L}^{\mu;\nu}(\pm\frac{1}{2})$ are the tensors with
helicity of the initial  lepton equal to $\pm 1/2$ and
\begin{eqnarray}
\label{lpm} {\cal L}^{\mu \nu}(+)&=& 2 (g^{\mu \nu} l \cdot q + 2
l^\mu l^\nu -
l^\mu q^\nu - l^\nu q^\mu),\nonumber \\
{\cal L}^{\mu \nu}(-)&=& 2 i \mu \epsilon^{\mu\nu\delta\rho}
q_{\delta} (s_{l})_{\rho}.
\end{eqnarray}
The hadronic tensor is given by a trace similar to the lepton case
(\ref{lept})
\begin{eqnarray}
\label{wtenz} W^{\alpha\alpha';\beta\beta'}(s_p)= \sum_{spin \;
s_f} \bar u(p',s_f)
 V_{pgg}^{\alpha\alpha'}(p,t,x_P,l) u(p,s_p) \nonumber\\ \bar u(p,s_p)
V_{pgg}^{\beta\beta'\,+}(p,t,x_P,l') u(p',s_f).
\end{eqnarray}

The spin--average and spin--dependent hadron tensors look like
\begin{equation}
\label{wpm} W^{\alpha\alpha';\beta\beta'}(\pm)=\frac{1}{2} (
W^{\alpha\alpha';\beta\beta'}(+s_p)
 \pm W^{\alpha\alpha';\beta\beta'}(-s_p)).
\end{equation}
Here $s_p$ is the spin vector  which determines the target
polarization. For the leading term of the spin- average structure
$W(+)$ for the ansatz (\ref{ver}) we find
\begin{equation}
\label{w+f} W^{\alpha\alpha';\beta\beta'}(+) = 16 p^{\alpha}
p^{\alpha'} p^{\beta}  p^{\beta'} ( |B|^2 + \frac{|t|}{m^2}
|K|^2).
\end{equation}
The  spin-dependent part of the hadron tensor is more complicated.
Its explicit form can be found in \cite{golostr}

The obtained equation for the spin-average tensor coincides in
form with the cross section of the proton off the spinless
particle (a meson, e.g.). Really, the meson--proton
helicity-non-flip and helicity-flip amplitudes can be written in
terms of the invariant functions $\tilde B$ and $\tilde K$ which
describe spin-non-flip and spin-flip effects
\begin{equation}
\label{fnf} F_{++}(s,t)= i s [\tilde B(t)] f(t);\;\; F_{+-}(s,t)=
i s \frac{\sqrt{|t|}}{m} \tilde K(t) f(t),
\end{equation}
where $f(t)$ is determined by the Pomeron coupling with meson. The
functions $\tilde B$ and $\tilde K$ are defined by integrals like
(\ref{spd}). The cross-section is written in the form
\begin{equation}
\label{mp_ds} \frac{d\sigma}{dt} \sim  [|\tilde B(t)|^2+
\frac{|t|}{m^2} |\tilde K(t)|^2] f(t)^2.
\end{equation}
The term proportional to $\tilde B$ represents the standard
Pomeron coupling that leads to the non-flip amplitude. The $\tilde
K$ function is the spin--dependent part of the Pomeron coupling
which produces in our case the transverse spin--flip effects in
the proton.

There are some models that provide  spin-flip effects which do not
vanish at high energies. In the model \cite{gol_mod}, the
amplitudes $K$ and $B$ have a phase shift caused by the soft
Pomeron rescattering effect.  The vector diquark in the diquark
model \cite{gol_kr} generates the  $K$ amplitude  which  is out of
phase with the Pomeron contribution to the amplitude $B$. The
models \cite{gol_mod,gol_kr} describe the experimental data
\cite{krish} on single spin transverse asymmetry $A_N$ quite well.
Thus, the weak energy dependence of spin asymmetries in exclusive
reactions is not now in contradiction with the experiment
\cite{gol_mod,akch}. The model \cite{gol_mod} predicts large
negative value of $A_N$ asymmetry near the diffraction minimum
which has a weak $s$ dependence in the RHIC energy range and $A_N$
is of about 10\% for $|t| \sim 3 \mbox{GeV}^2$ (Fig.1). Some other
model predictions for single-spin asymmetry at small momentum
transfer have been discussed in \cite{predazzi}. Thus, it should
be important to measure experimentally (PP2PP at RHIC) the
transverse asymmetry in the vicinity of diffraction minimum, which
is very sensitive to the imaginary part of the spin--flip $K$
amplitude. This can give direct information about the energy
dependence of the spin-flip amplitude $K$.

It has been found in \cite{gol_kr,gol_mod} that the ratio $|\tilde
K|/|\tilde B| \sim 0.1$ and has a weak energy dependence. We shall
use this value for our estimations of the asymmetry in diffractive
hadron production.
\begin{figure}
\centering \mbox{\epsfysize=50mm\epsffile{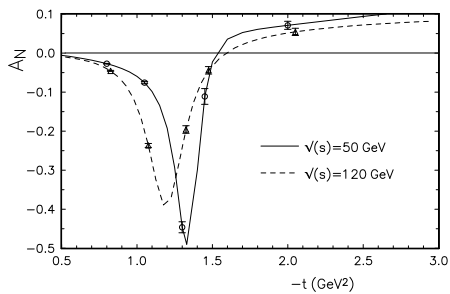}}
\phantom{.}\vspace{-0.5cm} \caption{Predictions of the model
\cite{gol_mod} for single-spin transverse asymmetry of the $pp$
scattering at RHIC energies
  \cite{akch}. Error bar indicates expected
 statistical errors for the PP2PP experiment at RHIC.} \label{F1}
\end{figure}

\section{Diffractive Vector Meson Leptoproduction}\label{sect4}
Vector meson production through the photon-two gluon fusion is
shown in Fig.2. We regard mainly the $J/\Psi$ meson production
where such effects are predominated. The $J/\Psi$ meson can be
considered as a $S$-wave system of heavy  quarks with the wave
function of the form
\begin{equation}\label{wf}
\Psi_V=g (/\hspace*{-0.20cm} k+m_q) \gamma_\mu
\end{equation}
where $k$ is the momentum of a quark, and $m_q$ is its mass. In
the nonrelativistic approximation the quarks have the same momenta
$k=V/2$ and the mass $m_q=m_J/2$. The transverse quark motion is
not considered. This means that the vector meson distribution
amplitude has a simple form $\delta(\tau-1/2)\delta(k_t^2)$. The
wave function (\ref{wf}) for nonzero mass $m_q$ produces both
amplitudes with a longitudinally and transversely polarized vector
meson. For the light meson production $m_q=0$,  and one must
consider the higher twist effects to calculate the amplitude with
transverse vector meson polarization (see e.g. \cite{mank}).
\begin{figure}
\centering \mbox{\epsfysize=30mm\epsffile{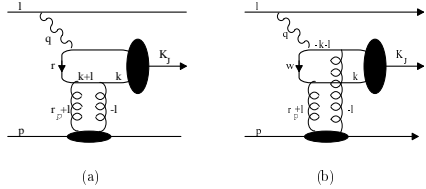}}
\phantom{.}\vspace{-0.5cm} \caption{Two-gluon contribution to
diffractive vector meson production.} \label{F2}
\end{figure}

The spin-average and spin-dependent cross-sections are defined as
\begin{equation}\label{ds0}
d \sigma(\pm) =\frac{1}{2} \left( d \sigma({\rightarrow}
{\Downarrow) \pm  d \sigma({\rightarrow} {\Uparrow}})\right).
\end{equation}
The cross section $d \sigma(\pm)$ can be written in the form
\begin{equation}\label{ds}
\frac{d\sigma^{\pm}}{dQ^2 dy dt}=\frac{|T^{\pm}|^2}{32 (2\pi)^3
 Q^4 s^2 y}.
\end{equation}
 For the spin-average  amplitude squared we find
\begin{equation}\label{t+}
|T^{+}|^2=  \frac{s^2\, N }{4 \bar Q^4}\,\left( (1+(1-y)^2) m_V^2
+ 2(1 -y) Q^2 \right) \left[ |\tilde B|^2+|\tilde K|^2
\frac{|t|}{m^2} \right].
\end{equation}
Here $\bar Q^2=(m_V^2+Q^2+|t|)/4$, and  $N$ is the normalization
factor. The term proportional to $(1+(1-y)^2) m_V^2$ represents
the contribution of the transversely  polarized photons. The $2(1
-y) Q^2$ term describes the effect of longitudinal photons (see
(\ref{slp})).

The function $\tilde B$ is determined by
\begin{eqnarray}\label{bt}
&&\tilde B= \bar Q^2 \int \frac{d^2l_\perp (l_\perp^2+\vec l_\perp
\vec r_\perp) B(t,l_\perp^2,x_P,...)} {(l_\perp^2+\lambda^2)((\vec
l_\perp+\vec \Delta)^2+\lambda^2)[l_\perp^2+\vec l_\perp \vec
r_\perp
+\bar Q^2]} \sim\nonumber\\
&& \int^{l_\perp^2<\bar Q^2}_0 \frac{d^2l_\perp (l_\perp^2+\vec
l_\perp \vec r_\perp) } {(l_\perp^2+\lambda^2)((\vec l_\perp+\vec
r_\perp)^2+\lambda^2)} B(t,l_\perp^2,x_P,...)={ {\cal
F}^g_{x_P}(x_P,t,\bar Q^2)}.
\end{eqnarray}
Due to the gauge invariance, the result in the Feynman gauge
should be equal to the corresponding result in the axial gauge
(section 2). This permits us to find  the connection with SPD in
(\ref{bt}) by a direct comparison of the cross section (\ref{ds})
with (\ref{s+i}). We can see that the $B(l_\perp^2,x_P,...)$
function represents the nonintegrated spin- average gluon
distribution.

The spin-dependent amplitude squared is found to be of the form
\begin{equation}\label{t-}
|T^{-}|^2=  \frac{\vec Q \vec S_\perp}{4 m}\; \frac{s |t| N}{4
\bar Q^4}\left(Q^2+ m_V^2+|t| \right) \frac{\tilde B \tilde K^*+
\tilde B^* \tilde K }{2} .
\end{equation}
It is determined by the interference between the $\tilde B$ and
$\tilde K$ amplitudes.
\section{Diffractive $Q \bar Q$ Photoproduction}
To study $Q \bar Q$ production we shall use the same two-gluon
model which should describe the cross sections at small $x<0.1$.
This contribution is shown in Fig. 3. The quark-antiquark
contribution, in addition to the $t$- channel gluons, is important
for light quark production at large $x$.
\begin{figure}
\centering \mbox{\epsfysize=20mm\epsffile{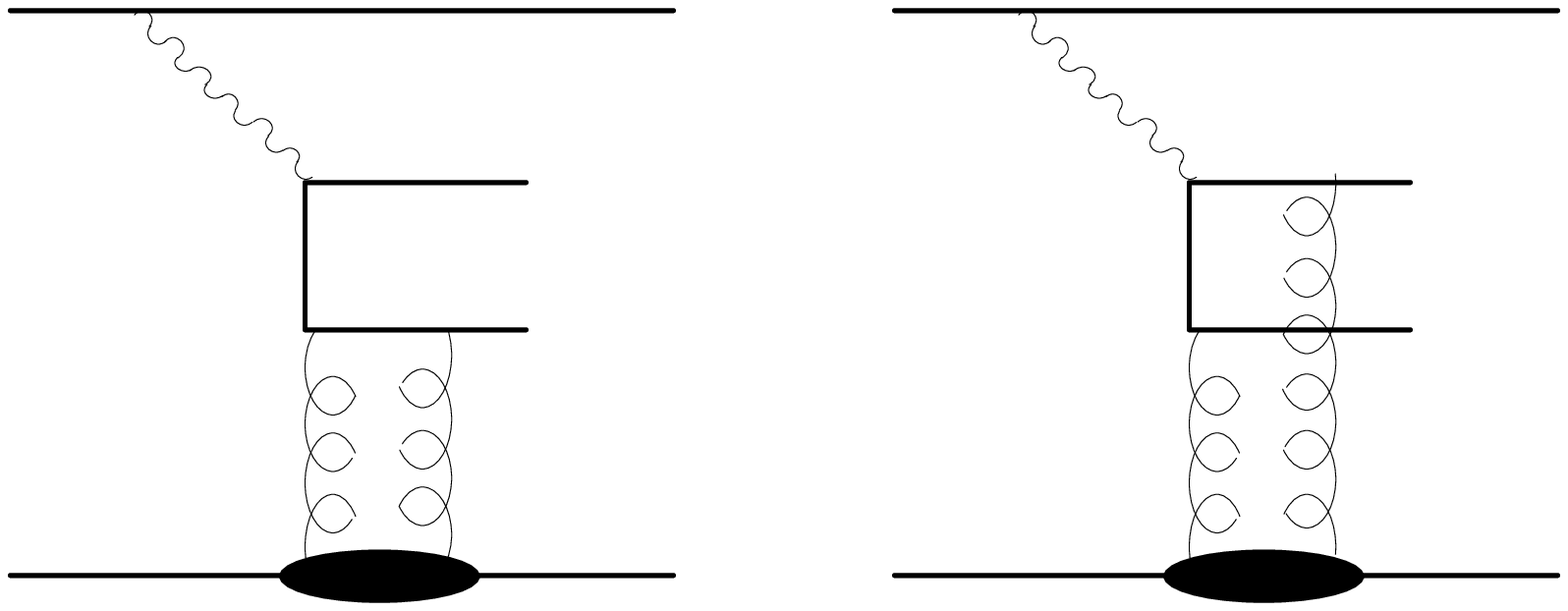}}
\caption{Two-gluon contribution to $Q \bar Q$ production}
\label{Fgg}
\end{figure}
To calculate  the cross sections, we  integrate the amplitudes
squared over the $Q \bar Q$ phase space $dN_{Q \bar
Q}=\Pi_f\frac{d^3p_f}{(2 \pi)^3 2 E_f}$ with the delta function
that reflects the momentum conservation. As a result, the
spin-average and spin-dependent cross section can be written in
the form
\begin{equation}
\label{sigma} \frac{d^5 \sigma(\pm)}{dQ^2 dy dx_p dt dk_\perp^2}=
\left(^{(2-2 y+y^2)} _{\hspace{3mm}(2-y)}\right)
 \frac{C(x_P,Q^2) \; N(\pm)}
{\sqrt{1-4(k_\perp^2+m_q^2)/M_X^2}}.
\end{equation}
Here $C(x_P,Q^2)$ is a normalization function which is common for
the spin average and spin dependent cross section; $N(\pm)$ is
determined by a sum of graphs integrated over the gluon momenta
$l$ and $l'$
\begin{equation}
\label{tpm} N(\pm) = \int \frac{ d^2 l_{\bot} d^2 l_{\bot}'
(l_\perp^2+\vec l_\perp \vec r_\perp)\; ((l'_\perp)^2+\vec
l'_\perp \vec r_\perp)\;D^{\pm}(t,Q^2,l_{\bot},l'_{\bot},\cdots)}
{(l_\perp^2+\lambda^2)((\vec l_\perp+\vec r_{\perp})^2+\lambda^2)
(l_\perp^{'2}+\lambda^2)((\vec l'_\perp+\vec
r_{\perp})^2+\lambda^2)}.
\end{equation}
The $D^{\pm}$ function is a hard part that contains the traces
over the quark loops of the graphs  convoluted with the spin
average and spin-dependent tensors.

The analytic forms of the graph contribution to the cross sections
will be written here in the limit $\beta \to 0$ for simplicity.
The  $D^{+}$ function has the following form \cite{golostr}:
\begin{equation}\label{d+}
D_I^{+}=\frac{Q^2 \left(|B|^2+|t|/m^2 |K|^2
\right)\left((k_\perp+r_\perp)^2+m_q^2 \right)}
{\left(k_\perp^2+m_q^2 \right)
\left((k_\perp-l_\perp)^2+m_q^2\right)
\left((k_\perp-l'_\perp)^2+m_q^2\right)}.
\end{equation}
This function contains a product of the off-mass-spell quark
propagators in the graphs. We can see that the quark virtuality
here is quite different as compared to the vector meson case. We
have no terms proportional to $Q^2$ (see (\ref{bt})). This will
change the scale in  gluon structure functions. Really, $l$ and
$l'$ are smaller than $k_\perp^2$ and the contribution of $D^p(+)$
to $N(+)$ is about
\begin{equation}\label{nn}
N^p(+) \sim \frac{\left(|\tilde B|^2+|t|/m^2 |\tilde K|^2 \right)
\left((k_\perp+r_\perp)^2+m_q^2 \right)}{\left(k_\perp^2+m_q^2
\right)^3}
\end{equation}
with
\begin{eqnarray}\label{bqq}
\tilde B \sim \int^{l_\perp^2<k_0^2}_0 \frac{d^2l_\perp
(l_\perp^2+\vec l_\perp \vec r_\perp) }
{(l_\perp^2+\lambda^2)((\vec l_\perp+\vec r_\perp)^2+\lambda^2)}
B(t,l_\perp^2,x_P,...) =  {\cal F}^g_{x_P}(x_P,t,k_0^2)\nonumber\\
\tilde K \sim \int^{l_\perp^2<k_0^2}_0 \frac{d^2l_\perp
(l_\perp^2+\vec l_\perp \vec r_\perp) }
{(l_\perp^2+\lambda^2)((\vec l_\perp+\vec r_\perp)^2+\lambda^2)}
K(t,l_\perp^2,x_P,...) =  {\cal K}^g_{x_P}(x_P,t,k_0^2),
\end{eqnarray}
where $k_0^2 \sim k_\perp^2+m_q^2$. We find that the structure
functions are determined by the same integrals as in (\ref{bt}),
but with  a different scale. The details of calculations can be
found in \cite{golostr}

The contribution of all graphs to the function $N(+)$ for an
arbitrary $\beta$ can be written as
\begin{equation}\label{np}
N(+)=\left(|\tilde B|^2+|t|/m^2 |\tilde K|^2 \right)
\Pi^{(+)}(t,k_\perp^2,Q^2).
\end{equation}
The function $\Pi^{(+)}$ will be calculated numerically.

The same analysis was done for the spin-dependent cross sections.
In addition to the term observed in (\ref{t-}) and proportional to
the scalar production $\vec Q \vec S_\perp$, the new term $\propto
\vec k_\perp \vec S_\perp$ appears in the spin-dependent part. As
a result, we find the following representation of the function
$N(-)$
\begin{eqnarray}\label{nm}
N(-)=\sqrt{\frac{|t|}{m^2}} \left(\tilde B \tilde K^*+\tilde B^*
\tilde K\right) [ \frac{(\vec Q \vec S_\perp)}{m}
\Pi^{(-)}_Q(t,k_\perp^2,Q^2) \nonumber\\
 +\frac{(\vec k_\perp \vec
S_\perp)}{m} \Pi^{(-)}_k(t,k_\perp^2,Q^2)].
\end{eqnarray}

\section{Numerical Results for Vector Meson production}
The spin-average cross section of the vector meson production at a
small momentum transfer is approximately proportional to the
$|\tilde B|^2$ function (\ref{t+}) which is connected with the
skewed gluon distribution.  The simple parameterization of the SPD
as a product of the form factor and the ordinary gluon
distribution will be used
\begin{equation}
\label{b_g} \tilde B(t,x_P, \bar Q^2) =F_B(t) \left( x_P
G(x_P,\bar Q^2)  \right).
\end{equation}
The form factor $F_B(t)$ is chosen as an electromagnetic form
factor of the proton. Such a simple choice can be justified by
that the Pomeron--proton vertex might be similar to the
photon--proton coupling
\begin{equation}
\label{fp} F_B(t) \sim F^{em}_p(t)=\frac{(4 m_p^2+2.8 |t|)}{(4
m_p^2+|t|)( 1+|t|/0.7GeV^2)^2}.
\end{equation}

The energy dependence of the cross sections is determined by the
Pomeron contribution to the gluon distribution function at small
$x$
\begin{equation}
\label{g_x} \left(x_P G(x_P,\bar Q^2) \right) \sim
\frac{const}{x_P^{\alpha_p(t)-1}} \sim \left( \frac{s
y}{m_J^2+Q^2+|t|} \right) ^{(\alpha_p(t)-1)}.
\end{equation}
Here $\alpha_p(t)$ is a Pomeron trajectory which is chosen in the
form
\begin{equation}\label{pom}
  \alpha_p(t)=1+\epsilon+ \alpha' t
\end{equation}
with $\epsilon=0.15$ and $ \alpha'=0$. These values are in
accordance with the fit of the diffractive $J/\Psi$ production by
ZEUS \cite{zeus97}.

We suppose that the ratio of spin--dependent and spin--average
densities  has a weak $x$ dependence and
\begin{equation}\label{ratio}
\frac{|\tilde K|}{|\tilde B|} \sim 0.1
\end{equation}
as in the case of elastic scattering. Our prediction for the cross
sections is shown in Fig.4.

The $A_{lT}$ asymmetry for vector meson production  is determined
by the ratio of cross sections determined in (\ref{t-},\ref{t+})
\begin{equation}
\label{asylt} A_{lT} \sim \frac{\vec Q \vec S_\perp}{4 m}\;
\frac{y x_P |t|}{ (1+(1-y)^2) m_V^2 + 2(1 -y) Q^2} \; \frac{\tilde
B \tilde K}{ |\tilde B|^2+|\tilde K|^2 |t|/m^2}.
\end{equation}
The expected asymmetry  for $J/\Psi$ production  at HERMES
energies  is shown in Fig.5 for the case when the transverse part
of the photon momentum is parallel to the target polarization
$S_\perp$.
 At the HERA energies, asymmetry will be extremely small.

 For a small momentum transfer,  this asymmetry can be approximated
as
\begin{equation}\label{alt}
 A_{LT} \sim C_g \frac{{\cal K}^g_\zeta(\zeta)}{{\cal F}^g_\zeta(\zeta)}
 \;\;\;\mbox{with} \;\zeta=x_P
\end{equation}
Simple estimations show that the coefficient $C_g(J/\Psi)$ at
HERMES energy for $y=0.5, |t|=1\mbox{GeV}^2, Q^2=5\mbox{GeV}^2$ is
quite small, about 0.007.
\\[4mm]
\begin{minipage}{6cm}
\epsfxsize=6.0cm \centerline{\epsfbox{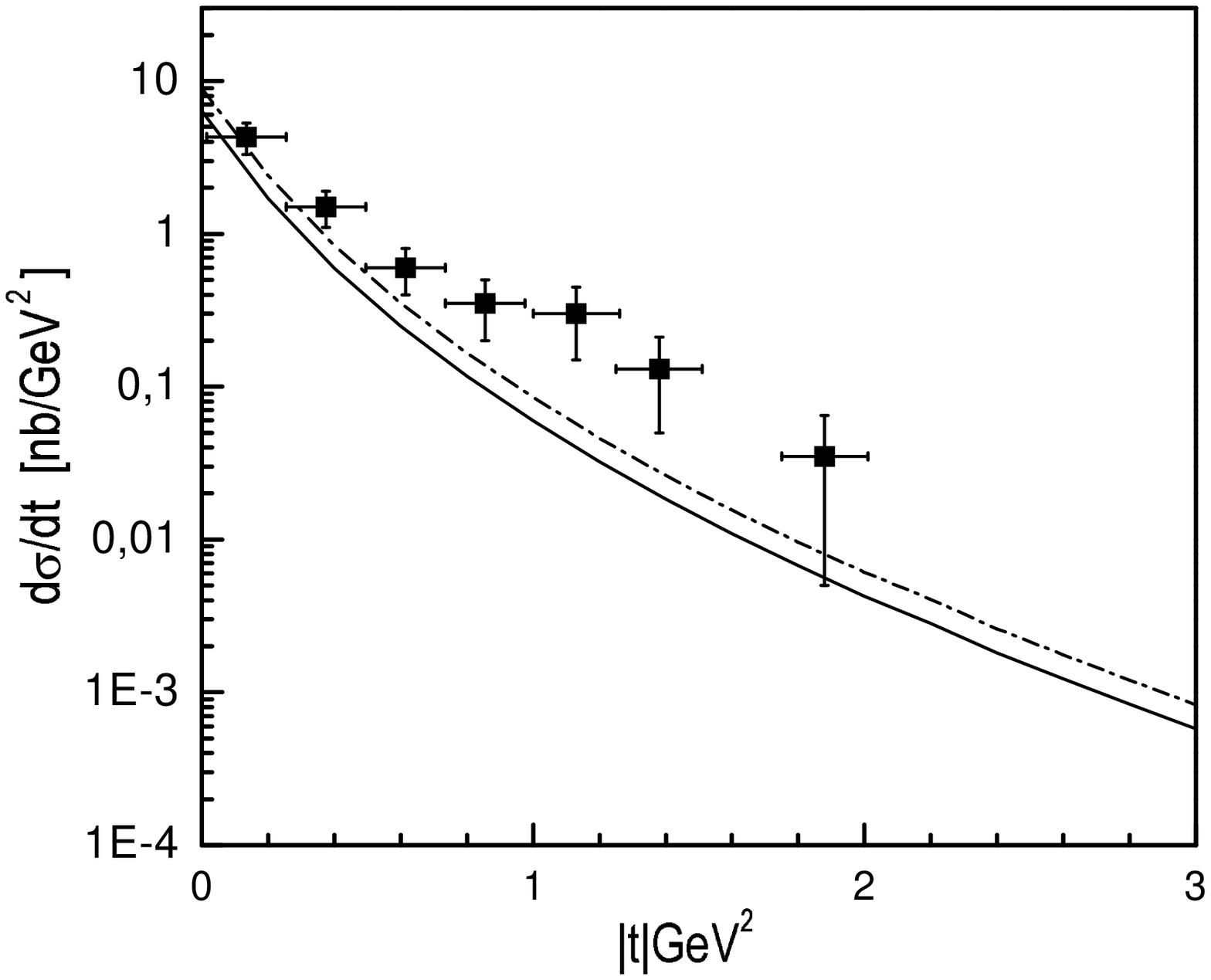}}
\end{minipage}
\begin{minipage}{0.14cm}
\phantom{aa}
\end{minipage}
\begin{minipage}{6cm}
\epsfxsize=6.0cm \centerline{\epsfbox{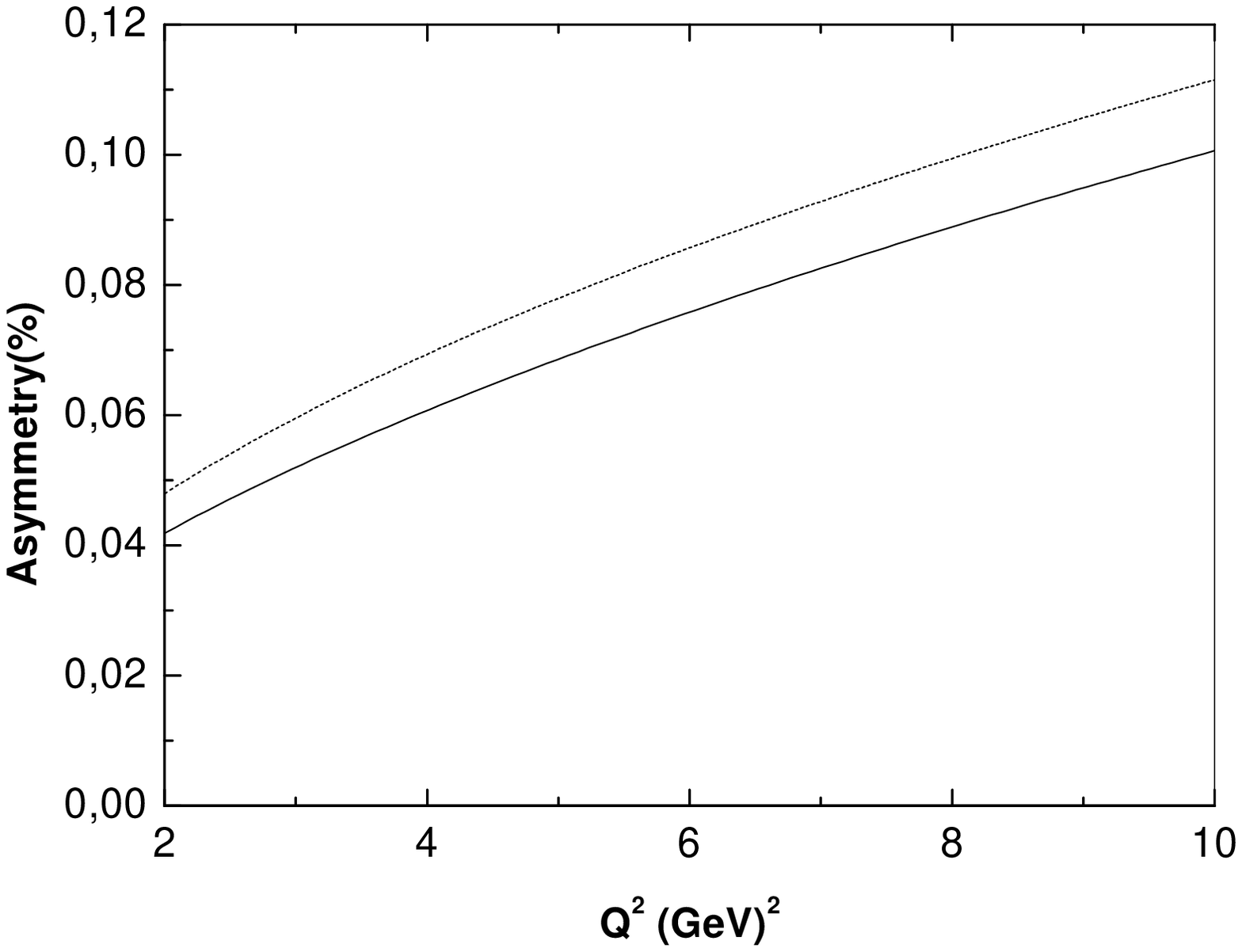}}
\end{minipage}
\\[.3cm]
\noindent
\begin{minipage}{6cm}
Fig. 4.~ {The differential cross section of  $J/\Psi$ production
at HERA energy: solid line -for $|\tilde K|/|\tilde B|=0$;
dot-dashed line -for $|\tilde K|/|\tilde B|=0.1$. Data are from
\cite{jpsi1}}.
\end{minipage}
\begin{minipage}{.15cm}
\phantom{aa}
\end{minipage}
\begin{minipage}{6cm}
Fig. 5.~ {The $A_{lT}$ asymmetry for vector meson production at
$\sqrt{s}=7 \mbox{GeV}$ ($y$=0.5, $|t|=1 \mbox{GeV}^2$): solid
line -for $J/\Psi$ production; dotted line -for $\rho$
production}.
\end{minipage}
\\[.3cm]

For light vector mesons, we use the same Eq. (\ref{asylt}).  The
model predicts a weak mass dependence of the gluon contribution to
the asymmetry, Fig.5. For the same kinematic variables, $C(\phi)
\sim C(\rho) \sim 0.008$. Note that this result was obtained for
the nonrelativistic meson wave function in the form
$\delta(\tau-1/2)\delta(k_t^2)$ which is not a good approximation
for light meson production. It is important to study a more
realistic wave function which  takes into consideration the
transverse quark degrees of freedom .
\section{Predictions for  $Q \bar Q$ Leptoproduction}\label{sect7}
 We shall now discuss our prediction for polarized diffractive $Q
\bar Q$ production.  In estimations of the asymmetry
$A_{lT}=\sigma(-)/\sigma(+)$ we shall use the same
parameterizations of SPD as in (\ref{b_g}) with the functions
determined in (\ref{fp}). As in the case of vector meson
production, the asymmetry is approximately proportional to the
ratio of polarized and spin--average gluon distribution functions
\begin{equation}\label{cltqq}
 A_{LT}^{Q \bar Q} \sim C^{Q \bar Q} \frac{{\cal K}^g_\zeta(\zeta)}
 {{\cal F}^g_\zeta(\zeta)}
 \;\;\;\mbox{with} \;\zeta=x_P\;\;\mbox{and}\; |\tilde K|/|\tilde B| \sim 0.1
\end{equation}

The spin-dependent contribution to the asymmetry which is
proportional to $\vec k_\perp \vec S_\perp$  will be analyzed for
the case when the transverse jet momentum $\vec k_\perp$ is
parallel to the target polarization $\vec S_\perp$. The asymmetry
is maximal in this case. To observe this contribution to
asymmetry, it is necessary to distinguish experimentally the quark
and antiquark jets. This can be realized presumably by the charge
of the leading particles in the jet which should be connected in
charge with the quark produced in the photon-gluon fusion. If we
do not separate events with $\vec k_\perp$ for the quark jet,
e.g., the resulting asymmetry will be equal to zero because the
transverse momentum of the quark and antiquark are equal and
opposite in sign.
\\[6mm]
\phantom{.}
\begin{minipage}{6cm}
\epsfxsize=5.9cm \centerline{\epsfbox{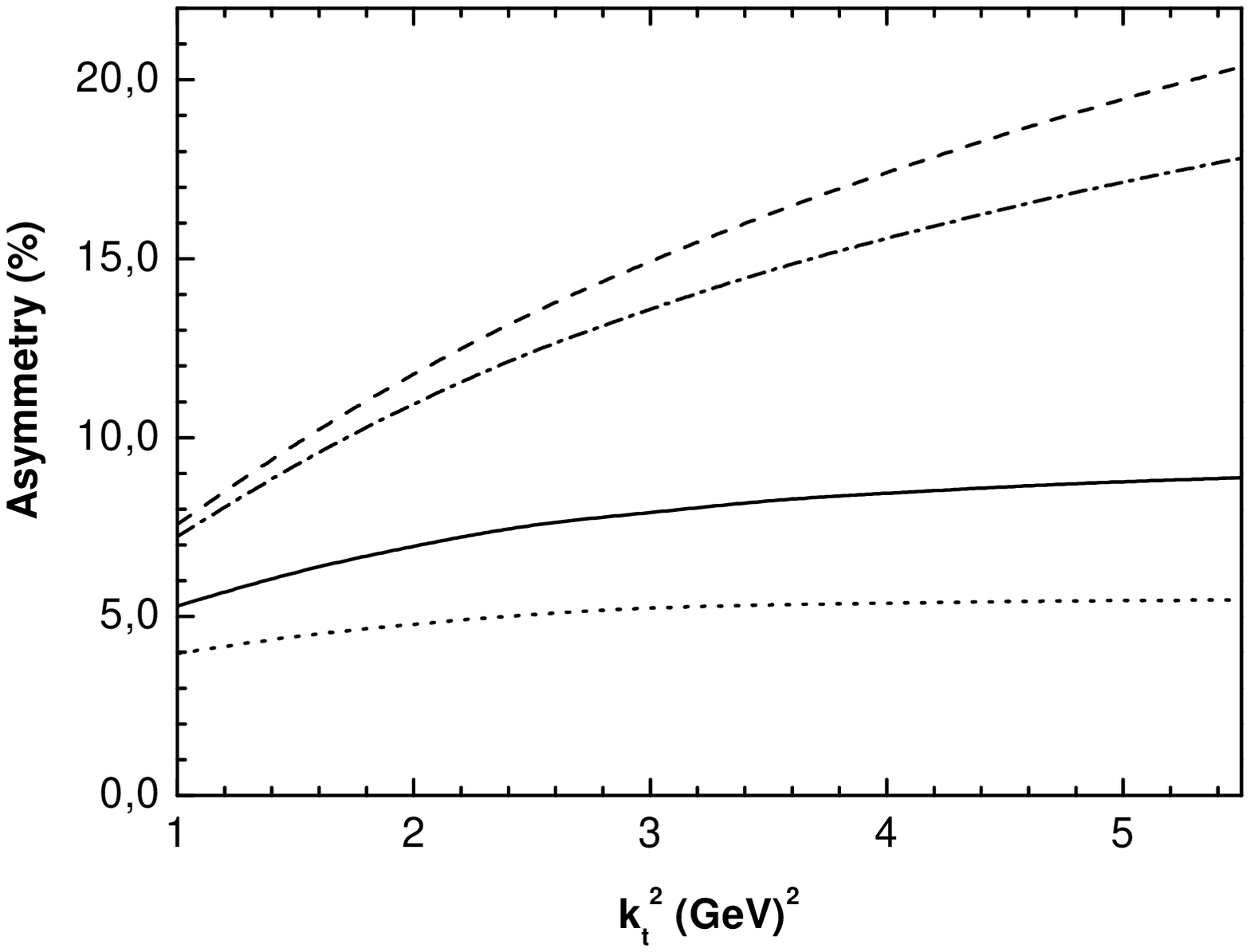}}
\end{minipage}
\begin{minipage}{0.14cm}
\phantom{aa}
\end{minipage}
\begin{minipage}{6cm}
\epsfxsize=6.0cm \centerline{\epsfbox{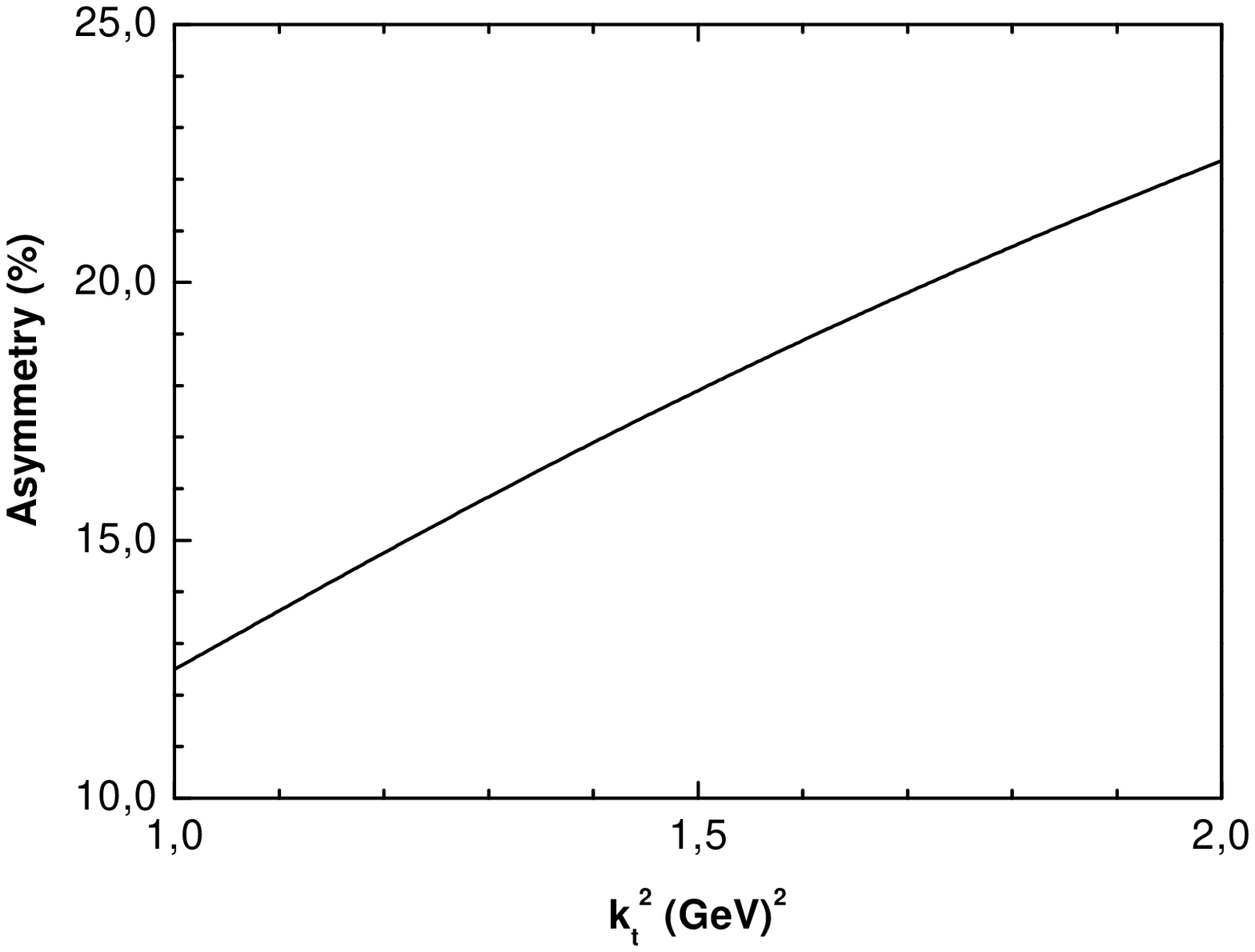}}
\end{minipage}
\\[.3cm]
\noindent
\begin{minipage}{6cm}
Fig. 6.~ {The $A^k_{lT}$ asymmetry in diffractive heavy $Q \bar Q$
production at $\sqrt{s}=20 \mbox{GeV}$ for $x_P=0.1$, $y=0.5$,
$|t|=0.3 \mbox{GeV}^2$: dotted line-for $Q^2=0.5 \mbox{GeV}^2$;
solid line-for $Q^2=1 \mbox{GeV}^2$; dot-dashed line-for $Q^2=5
\mbox{GeV}^2$; dashed line-for $Q^2=10 \mbox{GeV}^2$.}
\end{minipage}
\begin{minipage}{.15cm}
\phantom{aa}
\end{minipage}
\begin{minipage}{6cm}
Fig. 7.~ {The $A^k_{lT}$ asymmetry  in diffractive light $Q \bar
Q$ production for $Q^2=5\mbox{GeV}^2$, $x_P=0.1$, $y=0.5$,
$|t|=0.3(GeV)^2$ at $\sqrt{s}=7 \mbox{GeV}$.}
\end{minipage}
\\[.4cm]

The spin--dependent cross section  vanishes for $Q^2 \to 0$, while
the spin--average cross section is constant in this limit. As a
result, the $Q^2$ dependence of the asymmetry can be approximated
as  $A_{lT} \propto Q^2/(Q^2+Q^2_0)$ with $Q^2_0 \sim 1
\mbox{GeV}^2$. The predicted asymmetry for heavy $c \bar c$
production at COMPASS energies is shown in Fig.6. The asymmetry
for light quark production is  approximately of the same order of
magnitude. At the low energy $\sqrt{s} =7 \mbox{GeV}$ (HERMES) we
can work perturbatively only in a very limited region of $k^2$.
Really, $k_\perp^2$ should be large enough to have a large scale
$k_0^2$ in the process (\ref{bqq}). Otherwise, from (\ref{sigma}),
we have the restriction that $k^2\le M_X^2/4$. For quite large
$M_X^2 \sim (8-10) \mbox{GeV}^2 \sim M^2_{J/\Psi}$ we find that
$(k_\perp^2)_{max} \sim 2\mbox{GeV}^2$. The expected $A_{lT}$
asymmetry for light quark production at HERMES is shown in Fig.7.
The coefficient $C_k^{Q \bar Q}$ in (\ref{cltqq}) is quite large,
about 1.5 at the HERMES energy for $k_\perp^2=1.3\mbox{GeV}^2$,
$Q^2=5\mbox{GeV}^2$, $x_P=0.1$, $y=0.5$, and $|t|=0.3
\mbox{GeV}^2$. This shows a possibility of studying the polarized
gluon distribution ${\cal K}^g_\zeta(x)$ in the HERMES experiment.

The contribution to asymmetry $\propto \vec Q \vec S_\perp$ is
analyzed  for the case when the transverse jet momentum $\vec
Q_\perp$ is parallel to the target polarization $\vec S_\perp$ (a
maximal contribution to the asymmetry). The predicted $A^Q_{lT}$
asymmetry in diffractive heavy $Q \bar Q$ production at $\sqrt{s}
=20 \mbox{GeV}$ is shown in Fig.8. The $A^Q_{lT}$ asymmetry has a
strong mass dependence. For light quark production, this asymmetry
is not small for $Q^2 \sim (0.5-1) \mbox{GeV}^2$ and positive.
\\[5mm]
\begin{minipage}{6cm}
\epsfxsize=6.0cm \centerline{\epsfbox{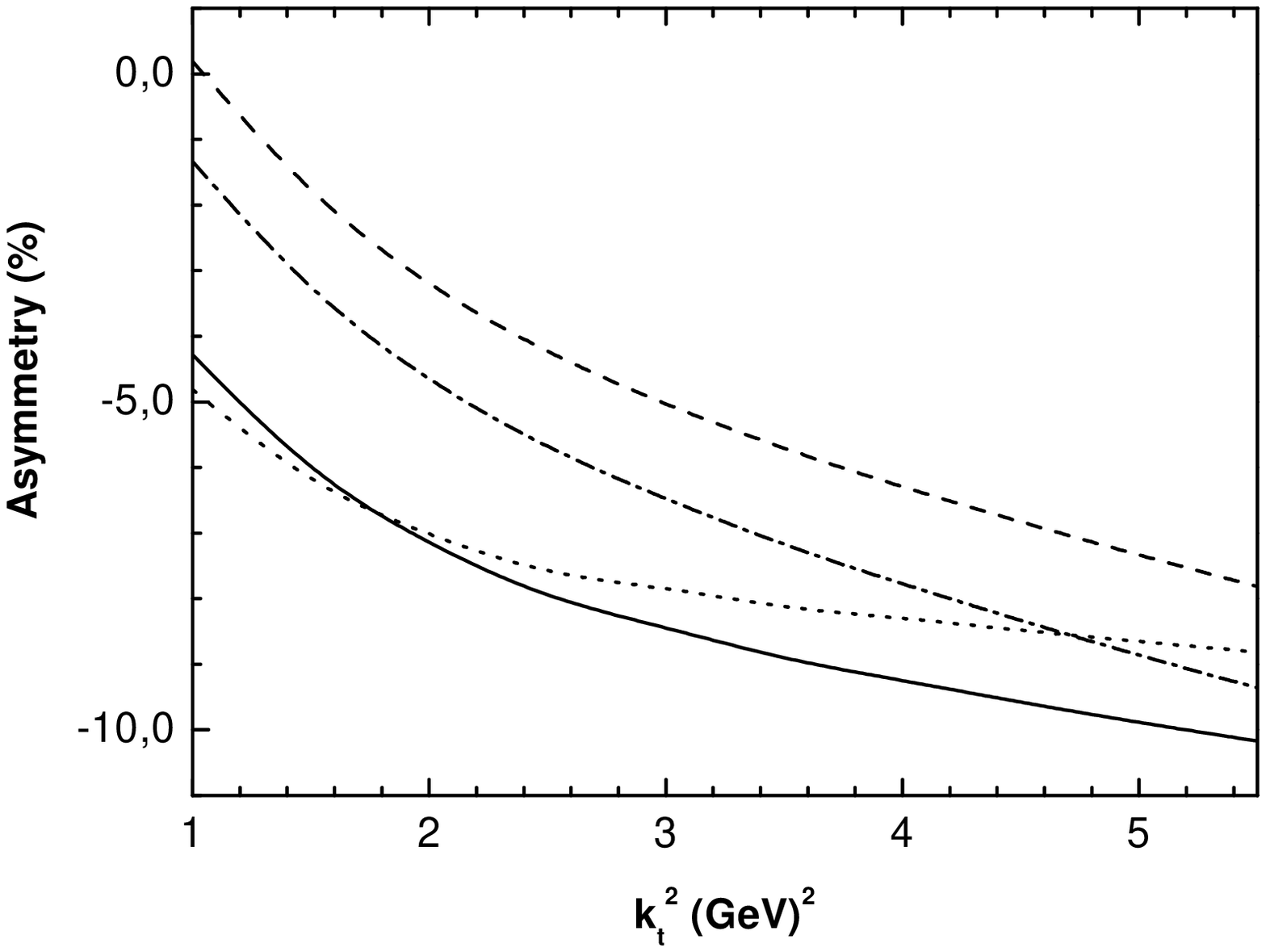}}
\end{minipage}
\begin{minipage}{0.14cm}
\phantom{aa}
\end{minipage}
\begin{minipage}{6cm}
\epsfxsize=6.0cm \centerline{\epsfbox{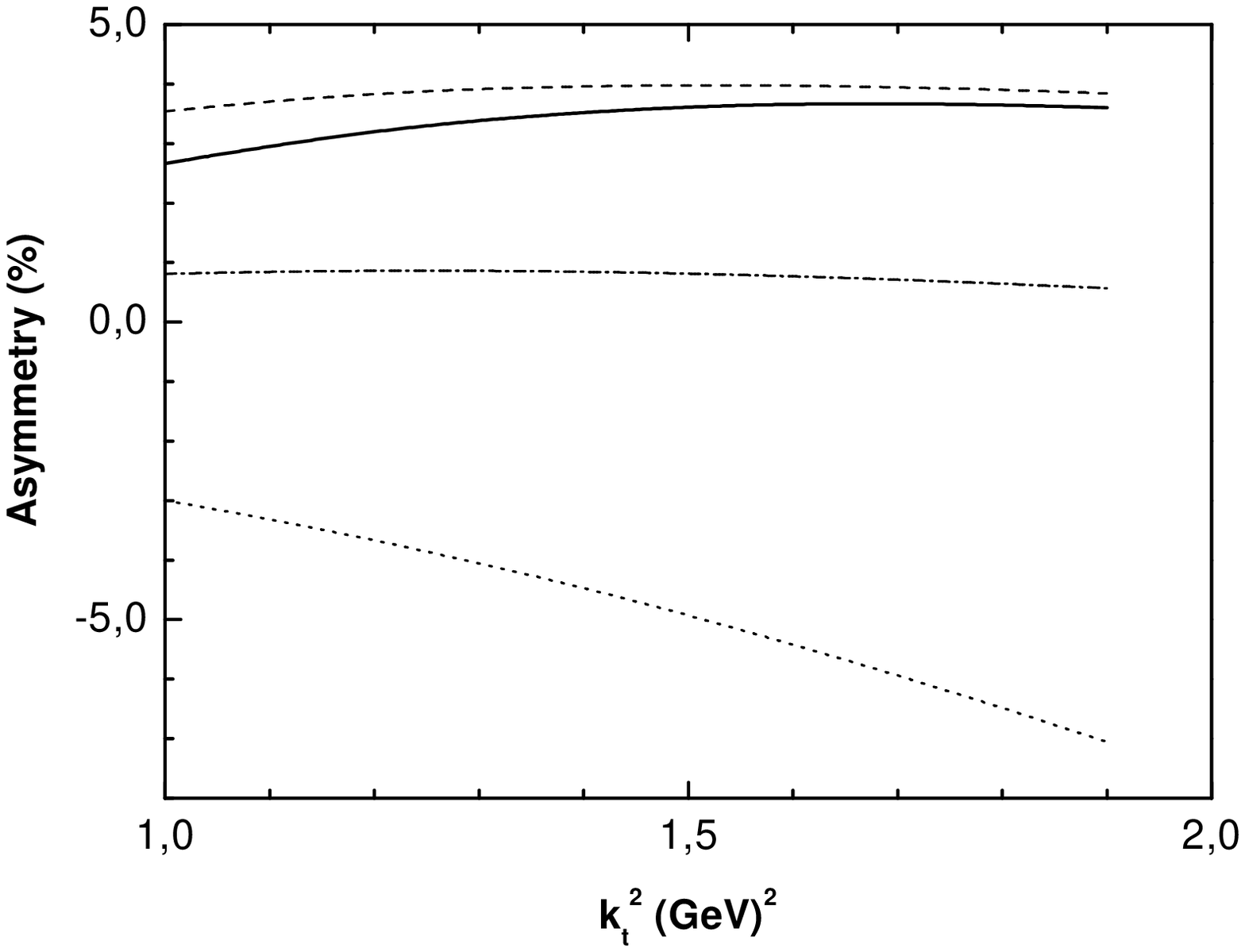}}
\end{minipage}
\\[.4cm]
\noindent
\begin{minipage}{6cm}
Fig. 8.~ {The $A^Q_{lT}$ asymmetry in diffractive heavy $Q \bar Q$
production at $\sqrt{s}=20 \mbox{GeV}$ for $x_P=0.1$, $y=0.5$,
$|t|=0.3 \mbox{GeV}^2$: dotted line-for $Q^2=0.5 \mbox{GeV}^2$;
solid line-for $Q^2=1 \mbox{GeV}^2$; dot-dashed line-for $Q^2=5
\mbox{GeV}^2$; dashed line-for $Q^2=10 \mbox{GeV}^2$.}
\end{minipage}
\begin{minipage}{.15cm}
\phantom{aa}
\end{minipage}
\begin{minipage}{6cm}
Fig. 9.~ {The The $A^Q_{lT}$ asymmetry  in diffractive light $Q
\bar Q$ production at  $\sqrt{s}=7 \mbox{GeV}$ for
$Q^2=5\mbox{GeV}^2$, $x_P=0.1$, $y=0.5$: dotted line -at $|t|=0.1
\mbox{GeV}^2$; dot-dashed line -at $|t|=0.3 \mbox{GeV}^2$; dashed
line -at $|t|=0.5 \mbox{GeV}^2$; solid line- integrated over $|t|$
asymmetry.}
\end{minipage}
\\[.3cm]

It is interesting to look for what we expect to observe for light
quark production at low energy $\sqrt{s} =7 \mbox{GeV}$. The
predicted asymmetry for different momentum transfers is shown in
Fig. 9. Note that in fixed--target experiments, it is usually
difficult to detect the final hadron and determine the momentum
transfer. In this case, it will be useful to have predictions for
the asymmetry integrated over momentum transfer
\begin{equation}\label{intasy}
\bar A^Q_{lT}= \frac{\int_{t_{min}}^{t_{max}}\,\sigma(-)\,dt}
{\int_{t_{min}}^{t_{max}}\,\sigma(+)\,dt}.
\end{equation}
We integrate cross sections from $t_{min} \sim (x_P m)^2 \sim 0$
up to $t_{max}=4 \mbox{GeV}^2$. The predicted integrated asymmetry
(see Fig. 9) is not small, about 3\%.

\section{Conclusion}\label{sect8}

In the present lecture, we analyze  diffractive hadron
leptoproduction at small $x$. Diffractive reactions are
predominated in this region by the gluon exchange. We consider two
approaches which are used in theoretical estimations of
diffractive processes. The first one is based on the factorization
of the scattering amplitude into the hard subprocess and soft
skewed gluon distribution. The other approach uses the two-gluon
exchange model where the cross sections of diffractive hadron
production are determined in terms of the leptonic and hadronic
tensors and the squared amplitude of hadron production through the
photon-two-gluon fusion. The hadronic tensor is expressed in terms
of the two--gluon couplings with the proton. The processes of
diffractive meson and $Q \bar Q$ production are dependent on the
same integrals over gluon transverse momentum which are related to
the gluon SPD ${\cal F}_\zeta(x)$ and ${\cal K}_\zeta(x)$
(\ref{bt},\ref{bqq}).

The diffractive hadron leptoproduction for a longitudinally
polarized lepton and a transversely polarized proton at high
energies has been studied within the two-gluon exchange model. The
$A_{lT}$ asymmetry is found to be proportional to the ratio of
structure functions $A_{lT}=C {\cal K}/{\cal F}$. This asymmetry
can be used to get information on the transverse distribution
${\cal K}^g_{x_P}(x_P,t)$ from experiment if the coefficient $C$
is not small. The corresponding coefficient for vector meson
production is expected to be quite small $C_g(J/\Psi) \sim 0.007$
in the HERMES energy range. It is difficult to expect experimental
study of such small asymmetry. However, this result was obtained
for a simple nonrelativistic form of the vector meson wave
function and can be used only for heavy meson production. The
light meson production is more complicated and asymmetry in this
case can be different. The transverse quark motion and higher
twist effects for transversely polarized $\rho$ meson should be
important here.

In the case of $Q \bar Q$ production we have additional transverse
variable $\vec k_\perp$. This produces, in addition to  the term
$\vec Q \vec S_\perp$ in the $A_{lT}$ asymmetry, the contribution
proportional to the scalar product $\vec k_\perp \vec S_\perp$
(\ref{nm}). These terms in the asymmetry have different kinematic
properties and can be studied independently. The term $\propto
\vec k_\perp \vec S_\perp$ has a large coefficient $C_k^{Q \bar
Q}$ that is predicted to be about 1. Such asymmetry might be an
excellent object to study transverse effects in the proton-- gluon
coupling. However, the experimental study of this asymmetry is not
so simple. To find nonzero asymmetry in this case, it is necessary
to distinguish quark and antiquark jets and to have a possibility
of studying the azimuthal asymmetry structure. The expected
$A_{lT}$ asymmetry for the term $\propto \vec Q \vec S_\perp$  is
not small too. The predicted coefficient $C_Q^{Q \bar Q}$ in this
case is about 0.3. The diffractive $Q \bar Q$ can be investigated
in a polarized proton- proton interaction too where the asymmetry
of the same order of magnitude as in the lepton-proton case was
predicted \cite{golos96}.

The results presented here should be applicable to the reactions
with heavy quarks. For processes with light quarks, our
predictions can be used in the small $x$ region ($x \le 0.1$ e.g.)
where the contribution of quark SPD is expected to be small. In
the region of not small $x\ge 0.1$  the polarized $u$ and $d$
quark SPD might be studied together with the gluon distribution in
the case of $\rho$ production. For the $\phi$ meson production,
the strange quark SPD might be analyzed. Such experiments can be
conducted at the future HERMES and COMPASS spectrometers for a
transversely polarized target. We conclude that important
information on the spin--dependent SPD ${\cal K}_\zeta(x)$ at
small $x$ can be obtained from the asymmetries in diffractive
hadron leptoproduction for longitudinally polarized lepton and
transversely polarized hadron targets.

These lectures were supported in part by the Russian Foundation of
Basic Research, Grant 00-02-16696 and by the Blokhintsev--Votruba
program.


\newpage
\begin{thebibliography}{9}
\bibitem{rad-j} A.V. Radyushkin: Phys.Rev. {\bf D56} (1997) 5524;
 X. Ji: Phys.Rev. {\bf D55} 7114 (1997).
\bibitem{zeus97} ZEUS Collab., J. Breitweg et al: Z. Phys,
{\bf C75} (1997) 215.
\bibitem{jpsi1} H1 Collaboration,  S. Aid et al: Nucl. Phys. {\bf B472} (1996)
 3.
\bibitem{h1_99} H1 Collab., C. Adloff et al: Eur. Phys. J. {\bf C10} (1999)  373.
\bibitem{dijet} ZEUS Collab., J. Breitweg et al: Eur. Phys. J. {\bf C5} (1998) 41;\\
               H1 Collab., C. Adloff et al: Eur. Phys. J. {\bf C6} (1999)  421.
\bibitem{dm} H1 Collab., C. Adloff et al: Eur. Phys. J. {\bf C103} (2000)
371;\\
ZEUS Collab., J. Breitweg et al: Eur. Phys. J. {\bf C12} (2000)
393.
\bibitem{hermes} HERMES Collab., A. Airapetian et al: Phys. Lett. {\bf B513} (2001) 301.
\bibitem{rys} M.G. Ryskin: Z. Phys. {\bf C57} (1993) 89.
\bibitem{J-Psi} S.J. Brodsky at al: Phys. Rev. {\bf D50} (1994)
3134;\\
L. Frankfurt, W. Koepf, M. Strikman:  Phys. Rev. {\bf D57} (1998)
512.
\bibitem{clebaux}B. Clerbaux: {\it Elastic production of Vector Mesons at HERA:
  study of the scale of the interaction and measurement of the helicity amplitudes}.
  E-print: hep-ph/9908519.
\bibitem{rys1} M.G. Ryskin, R.G. Roberts, A.D. Martin, E.M. Levin:
     Z. Phys.  {\bf C76} (1997) 231.
\bibitem{cud} J.L. Cudell, I. Royen: Nucl. Phys. {\bf B545} (1999) 505.
\bibitem{ivan} D.Y. Ivanov, R. Kirshner:  Phys. Rev. {\bf D58} (1998) 114026.
\bibitem{mank98} L.\ Mankiewicz, G. Piller, T. Weigl:  Eur. Phys. J. {\bf C5} (1998) 119.
\bibitem{hua_kr} H.W. Huang, P. Kroll: Eur. Phys. J. {\bf C17} (2000)  423.
\bibitem{mank} M.\ V\"anttinen, L.\ Mankiewicz: Phys.Lett.  {\bf B434} (1998) 141.
\bibitem{die95} M.Diehl: Z. Phys. {\bf C66} (1995) 181.
\bibitem{bart96} J. Bartels, C. Ewerz, H. Lotter, M.W\"usthoff:
                Phys. Lett. {\bf B386} (1996) 389.
\bibitem{ryskin97}  E.M. Levin, A.D. Martin, M.G. Ryskin, T. Teubner:
            Z. Phys.  {\bf C74} (1997) 671.
\bibitem{schaef} B. Lehmann-Dronke, M. Maul, S. Schaefer, E.Stein, A. Sch\"afer:
       Phys.Lett. {\bf B457} (1999) 207.
\bibitem{mank00} L.\ Mankiewicz, G. Piller: Phys. Rev. {\bf D61} (2000) 074013.
\bibitem{grib77} L.V. Gribov, E.M. Levin,  M.G. Ryskin: Phys.
             Rept. {\bf 100} (1983) 151.
\bibitem{golostr} S.V.\ Goloskokov: Euro.\ Phys.\ J. {\bf C24}   (2002) 413.
\bibitem{gol_mod}  S.V. Goloskokov, S.P. Kuleshov, O.V. Selyugin:
           Z. Phys. {\bf C50} (1991)  455.
\bibitem{gol_kr} S.V.\ Goloskokov, P.\ Kroll: Phys.\ Rev.
          {\bf D60} (1999) 014019.
\bibitem{krish} G. Fidecaro  et al: Phys. Lett.  {\bf B76} (1978)
369;  {\bf B105} (1981) 309.
\bibitem{akch}  N. Akchurin, S.V. Goloskokov, O.V. Selyugin:
Int.J.Mod.Phys. {\bf A14} (1999) 253.
\bibitem{predazzi}  E. Predazzi: lectures at this
conference; \hspace{1mm}  A.F. Martini, E. Predazzi: {\it
Diffractive effects in spin-flip pp amplitudes and predictions for
relativistic energies.} E-print: hep-ph/0209027.
\bibitem{golos96} S.V.Goloskokov: Phys.Rev. {\bf D53} (1996) 5995.
\end{thebibliography}
\end{document}